# Continuum Electrostatics Approaches to Calculating p$K_a$s and E$_m$s in Proteins


*MR Gunner and Nathan A. Baker*

Physics Department, City College of New York in the City University of New York

160 Convent Avenue, New York NY 10031.  Email: mgunner@ccny.cuny.edu

Computational and Statistical Analytics Division, Pacific Northwest National Laboratory;

Division of Applied Mathematics, Brown University.  Email: nathan.baker@pnnl.gov






Key words: Dielectric constant, Electrostatics, Monte Carlo, Molecular Simulation, $pK_a$,

Poisson Boltzmann,





## Abstract

Proteins change their charge state through protonation and redox reactions as well as through binding charged ligands. The free energy of these reactions are dominated by solvation and electrostatic energies and modulated by protein conformational relaxation in response to the ionization state changes. Although computational methods for calculating these interactions can provide very powerful tools for predicting protein charge states, they include several critical approximations of which users should be aware. This chapter discusses the strengths, weaknesses, and approximations of popular computational methods for predicting charge states and understanding their underlying electrostatic interactions. The goal of this chapter is to inform users about applications and potential caveats of these methods as well as outline directions for future theoretical and computational research.





# **Introduction**

Methods that use continuum electrostatics have been developed to calculate the energies of protein charge states as they change through processes such as residue protonation, redox chemistry, or ion binding. While only a subset of amino acids are titratable, they play key roles in protein function (Bartlett, Porter, Borkakoti & Thornton, 2002). The model $pK_a$ of isolated amino acid in aqueous solution (Richarz & Wüthrich, 1975), can be used to calculate the probability that an isolated residue is charged at a given pH. Aspartate (Asp), glutamate (Glu), arginine (Arg), and lysine (Lys) comprise approximately 25% of average proteins and their $pK_{a,sol}$ values favor their ionization at physiological pHs (Kim, Mao & Gunner, 2005). The termini of amino acid chains also have model $pK_a$ values that cause them to frequently be ionized at physiological pH. Isolated histidine (His) has a $pK_a$ value near 7, which makes it easy to titrate at physiological pH values. It not surprising that His is highly enriched in active sites (Holliday, Almonacid, Mitchell & Thornton, 2007). Cysteine (Cys) (Go & Jones, 2013) and tyrosine (Tyr) (Styring, Sjoholm & Mamedov, 2012) are acids with higher model $pK_a$ values and are therefore less frequently ionized; however, these residues can play important functional roles as proton donors and as redox active sites. The residues in the active sites of proteins are often made of clusters of residues with linked protonation equilibria, leading to "non-ideal" titration curves (Ondrechen, Clifton & Ringe, 2001). A remarkable number of the mutations that lead to cancer involve protonatable residues (Webb, Chimenti, Jacobson & Barber, 2011). Although nucleic acids (Wong & Pollack, 2010) and phospholipid membranes (Argudo, Bethel, Marcoline &





Grabe, 2016) also have titratable groups and strong electrostatic interactions, this chapter will focus on proteins.

The ionization states of small molecules are important to their functions as substrates, cofactors, and control factors. Many protein ligands are charged with ionization states that can change during the binding process or enzymatic reactions (Schindler, Bornmann, Pellicena, Miller, Clarkson & Kuriyan, 2000; Dissanayake, Swails, Harris, Roitberg & York, 2015; Lee, Miller & Brooks, 2016). Cofactors such as NAD or FAD have charged groups such as phosphates that do not participate in reactions but must be bound to the protein for enzyme catalysis. Metal ions are often used by proteins to enhance stability such as in Zn fingers and as participants in redox reactions (Williams, 1997). Biological processes often occur at salt concentrations of 150 mM (or higher (Bowers & Wiegel, 2011)) such that all biomolecules are surrounded by a bath of small ions. The resulting ion cloud interacts with the protein in several ways, including salt-specific protein binding, electrostatic screening, and changing the thermodynamic activity of the protein in solution (Record, Anderson & Lohman, 1978; Grochowski & Trylska, 2008)

Charged groups have very favorable interactions with water that strongly influence their behavior in aqueous solutions (Warshel & Russell, 1984; Ren et al., 2012). They are often found on the on protein exterior surfaces maximizing their interaction with water while ensuring protein solubility and influencing interactions with other biomolecules. Supercharged proteins with total charges in excess of ±30 e are now used in protein design to prevent aggregation (Lawrence, Phillips & Liu, 2007). However, an important minority of charges are found within proteins, where they play functional roles.





Protein interiors are not simple hydrophobic environments and thus can tolerate internal charges through favorable electrostatic interactions (Spassov, Ladenstein & Karshikoff, 1997; Kim et al., 2005). Interaction with other buried charges can form stabilizing ion pairs. Additionally, polar and polarizable groups are also present in protein interiors: the amide backbone dipole moment is larger than water's (Gunner, Saleh, Cross, ud-Doula & Wise, 2000), many amino acid side chains that are polar or polarizable, and many crystal structures show water molecules and ions in protein interiors (Nayal & Di Cera, 1994, 1996; Makarov, Pettitt & Feig, 2002) One of the goals of the electrostatic calculation methods described in this chapter is to quantitatively understand the nature of these electrostatic interactions.

This review will describe current methods for computing the charge states of residues and ligands as a function of pH, $E_h$, or solution salt concentrations. The key reactions are thus:

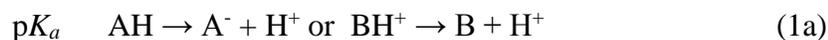

$pK_a$      $AH \rightarrow A^- + H^+$ or $BH^+ \rightarrow B + H^+$            (1a)

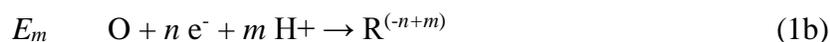

$E_m$      $O + n\,e^- + m\,H+ \rightarrow R^{(-n+m)}$            (1b)

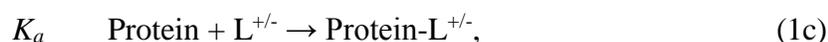

$K_a$      $Protein + L^{+/-} \rightarrow Protein\text{-}L^{+/-}$,            (1c)

where the species include acids (A), bases (B), ligands (L), oxidized species (O), and reduced species (R). Redox reactions are characterized by the number of electrons ($n$) and hydrogens ($m$) transferred. Ligand binding may also be accompanied by changes in protonation of the protein or ligand (Lee et al., 2016). Computational methods for modeling these reactions attempt to predict how the energetics of proton ($pK_a$), electron ($E_m$), or ligand ($K_a$) change as a function of environment (e.g., protein interior vs. solution). A wide range of experimental data are available for testing the predicted $pK_a$ (Stanton & Houk, 2008; Gosink, Hogan, Pulsipher & Baker, 2014),





$E_m$ (Reedy & Gibney, 2004) and $K_d$ (Gilson, Liu, Baitaluk, Nicola, Hwang & Chong, 2016) values. The goal of matching specific numerical values creates a high bar for testing these calculation methodologies.

The p$K_a$ value is the solution pH where the activities of $A^-$ and AH (or B and $BH^+$) are equal. However, proton/electron/ion binding affinities depend on the pH-dependent ionization states of other protein residues, thus making the proton affinity and thus the *in situ pK$_a$* pH-dependent. Therefore, a single p$K_a$ is often insufficient for characterizing the behavior of a titratable residue. Titration curves, describing the charge state as a function of pH can provide valuable information about the energetics influencing protein charge regulation (H. Webb et al., 2011). In addition, as reactions often occur far from the $pK_a$ of the reactant, it is often important to determine the proton affinity at physiological pH (Goyal, Lu, Yang, Gunner & Cui, 2013). Insight into protein electrostatics is ideally obtained through the (favorable) comparison of experimental and calculated titration curves together with the microscopic information (e.g., electrostatic potentials, hydrogen bonding networks, etc.) obtained from computational methods (Nielsen, Gunner & Garcia-Moreno, 2011).

The calculation of the free energy of a group of charges in a protein or other macromolecule by continuum methods has been reviewed extensively (Gunner & Alexov, 2000; Bashford, 2004; Garcia-Moreno & Fitch, 2004; Warshel, Sharma, Kato & Parson, 2006; Alexov et al., 2011). Thus, rather than providing a detailed methods review, we will highlight the information and choices that guide continuum electrostatics calculation and discuss emerging strategies for improving these methods.





# **Biomolecular structure and flexibility**

Structures are a required starting point for contemporary continuum electrostatic calculations (Berman et al., 2000). Charge state calculations are sensitive to structural details so care should be taken to access the structural quality using tools now found associated with all structures in the PDB database. However, rigid, single structures are inadequate for accurate charge state calculations due to the importance of changes in flexibility and conformation that occur upon introduction of new charges into a protein.

One of the key choices in charge state modeling involves the degrees of freedom (DOFs) included in the model. In the simplest case of rigid molecules, the only DOFs are the protonation or redox states or the ligand binding state. Because the protein will move in response to changes in the protonation, redox, or binding states, sampling DOFs for multiple structural "conformers" available to protein or ligand allows more a more "physical" analysis of the process. Thus, continuum electrostatics simulations balance implicit DOFs which, as described below, are approximated by the dielectric constant of the protein ($\varepsilon_{\text{solute}}$) and explicit degrees of freedom.

A protein microstate is a defined choice for each element that has any DOF. Each microstate has an associated energy that is used to generate the thermodynamic averages for titration curves, p$K_a$s, $E_m$s, binding probabilities, etc. The energy $G_\alpha$ of a microstate $\alpha$ can be written as a sum of contributions, which is implicitly summed over each group with DOF:

$$G_\alpha = m_\alpha \mu_\alpha + U_\alpha^{\text{MM}} + U_\alpha^{\text{elec}} + \Delta G_\alpha^{\text{p}} + \Delta G_\alpha^{\text{np}} \qquad (2)$$

where $m_\alpha \mu_\alpha$ is the free energy of $m_\alpha$ bound species with chemical potential $\mu_\alpha$, $U_\alpha^{\text{MM}}$ is the non-





electrostatic molecular mechanical energy, $U_\alpha^{\text{elec}}$ is the electrostatic energy, the $U_\alpha$ terms depend explicitly on the state of the other residues in the microstate, $\Delta G_\alpha^{\text{p}}$ is the polar solvation energy, and $\Delta G_\alpha^{\text{np}}$ is the nonpolar solvation energy. The chemical potential $\mu_\alpha$ varies for the quantity of interest; for protonation, it can be written as:

$$\mu_\alpha = \pm k_B T \log(10) \left( pH - pK_{a,\alpha} \right) \tag{3}$$

where $k_B T \approx 2.5$ kJ/mol (0.43 p$K_a$ units) is the thermal energy at room temperature and $\log(10) \approx 2.3$. Most energies represent energy differences between the microstate in the protein interior and a reference state in solution. The quantity $pK_{a,\alpha}$ is the model value for the residue in solution: the positive form of the expression is used for acidic sites, and the negative form is used for basic sites. Given this reference value, the other terms in the equation represent an effective shift to the model $pK_{a,\alpha}$ to account for the influence of the protein environment.

The probability of a microstate $\beta$ is given by the ensemble average

$$p_\beta = \frac{\sum_{\alpha \neq \beta} e^{-G_\alpha / k_B T}}{\sum_\alpha e^{-G_\alpha / k_B T}}. \tag{4}$$

If $n$ groups each sample 2 protonation states, then there are $2^n$ microstates. If residue $i$ has $m_i$ protonation or steric conformers there are $\prod_i m_i$ microstates where the product runs over all residues with DOF. The high dimensionality of this sum makes it impractical to evaluate for most protein systems. Instead of direct evaluation, $\rho_\beta$ is often calculated through limited conformational sampling; e.g., via Monte Carlo (MC) simulations (Song, Mao & Gunner, 2009; Polydorides & Simonson, 2013). Conformational degrees of freedom can range from sampling side chain rotameric states to relatively inexpensive optimization of steric clashes (Song, 2011)





and simple enumeration of different tautomeric forms for the hydrogen position on protonated side chains.

Allowing only dipolar groups to reorient and sample multiple rotameric and tautomer states has significant advantages (Nielsen & Vriend, 2001). Modifying these positions, remodels the hydrogen bond network in response to charge changes, which can provide a significant energetic stabilization of titration events. Note that this form of limited sampling can require *ad hoc* entropy corrections to compensate for larger numbers of neutral state tautomers or conformers (Song et al., 2009).

One significant approximation in conformational sampling involves the treatment of the intramolecular interactions, which use force fields with only self and pairwise energetics to greatly improve the efficiency when evaluating microstate energies: the $U_\alpha$ terms include only pairwise additive energetics between two groups and are independent of the state of any third group. Evaluation of all pairwise interactions yields an energy matrix of dimension $m^2$ for $m$ conformers. This pairwise decomposition is possible for less accurate non-polarizable force fields and algorithms that sample proton positions and side chain rotameric states. However, more recent polarizable force fields and larger collective protein motions such as backbone displacements generally cannot be represented in this pairwise form. However, while most methods that utilize Monte Carlo sampling make this approximation, it should be recognized that it misses motions that are likely to be important (Richman, Majumdar & Garcia-Moreno, 2014).

Monte Carlo methods can be used to incorporate side chain conformer sampling on a rigid protein backbone (Rabenstein, Ullmann & Knapp, 1998; Song et al., 2009; Polydorides &





Simonson, 2013).  Such sampling attempts to explicitly evaluate the ensemble average described above and thus incorporates a significant amount of side chain response to charge state changes. However, adding conformational degrees of freedom requires new approximations.  In particular, the shape of the protein can change when sampling different side chain or backbone conformations.  Calculation costs can increase dramatically if the shape is each microstate is calculated explicitly.  To maintain the cost for calculating the interaction energies of $O(m^2)$ often relies on all conformers being present when the continuum electrostatic pairwise interactions between conformers are determined.  This can exaggerate the low dielectric space of the protein. Some early approaches scaled the electrostatic interactions by an empirical screening function (Georgescu, Alexov & Gunner, 2002).  Newer methods correct for dielectric boundary errors due to excess conformers by using information obtained from a small number of calculations with an exact boundary (Song et al., 2009).

Molecular dynamics calculations can also be used to provide conformations in different protonation states. Given a particular titratable site, two sets of simulations are performed:  one with the charged state of the site and a second with the neutral state.  For sufficiently small energetic differences between the two protonation states (i.e., when linear response theory is valid), these ensembles will substantially overlap and the titration probability can be calculated via simple ensemble averages (Sham, Chu & Warshel, 1997).  Molecular dynamics-based linear response approaches have two key limitations.  The first is the computational expense of running $O(2^n)$ molecular dynamics simulations to sample the $n$ distinct neutral and ionized charge states. Rational choices can help pick consequential protonation states to sample (Witham et al., 2011; Meyer & Knapp, 2015). The second limitation is the underlying linear response assumption





requiring the energy difference between neutral and ionized states is small -- which is often not true for the important titration events in protein systems (Di Russo, Marti & Roitberg, 2014).

Molecular dynamics (MD) simulations can also be performed in open constant-pH ensembles. Unlike the fixed charge state simulations described above, constant-pH methods allow charged sites to exchange protons with the surrounding solution based on the pH of the bulk media, the model $pK_a$ of the site, and the energetics of the conformational ensemble. One class of methods performs MD for 10s of fs followed by a continuum electrostatics $pK_a$ calculation as described above to modify the protonation states in the MD trajectory (Baptista, Martel & Peterson, 1997; Swails, York & Roitberg, 2014; Lee, Miller, Damjanovic & Brooks, 2015). Alternatively, protonation states can be changed continuously via λ dynamics (Khandogin & Brooks, 2005; Goh, Hulbert, Zhou & Brooks, 2014). The primary hurdle to adoption of such continuous-pH MD methods is the difficulty of reaching convergence of the simulations. The use of pH replica exchange has led to significant improvements, but these methods are still not routine for the study of large proteins.

## **Solvent models or: How I learned to stop worrying and love the dielectric coefficient**

Most methods use continuum models of solvation behavior to incorporate the effects of solvent on charging energetics because of the computation effort associated with explicit descriptions of water molecules. The simplest continuum model for electrostatics represents the solvent as a dielectric material, usually with a dielectric coefficient of approximately 80 to





represent water. The Poisson equation (Nicholls & Honig, 1991; Baker, 2004) describes polar

solvation (electrostatic) energies within this dielectric approximation.  This equation is can be

combined with a nonpolar solvation term that describes the non-electrostatic contributions from

the solvent when conformers with significantly different surface exposure are sampled (Song et

al., 2009).  For calculations of ligand affinity and for ligand partition coefficients, continuum

models generally include a shape-related contribution, to describe the work associated with

inserting the uncharged solute into water, and a Lennard-Jones-like term to describe the weak

solute-solvent dispersive interactions (Lee et al., 2016).  Although small, such weak dispersive

forces play an important role in protein solvation and in titration state calculations (Levy, Zhang,

Gallicchio & Felts, 2003; Wagoner & Baker, 2004; Song et al., 2009).  Popular models for the

cavity term generally contain a term that scales as the area of the molecule times the surface

tension of the solution and often also include a term that multiples solution (hard sphere)

pressure with the volume of the solute (Wagoner & Baker, 2006).

Use of the Poisson equation – or other related continuum models – assumes that all

polarization in the system (molecule and solvent) is linear, local, and time-independent.  Linear

response implies that, no matter how large the electric field, the system will polarize in a

proportional manner.  However, given the finite density and polarizability of water and

molecular solutes, this assumption is clearly violated at high charge densities and field strengths,

such as found near nucleic acids (Lipfert, Doniach, Das & Herschlag, 2014).  Local response

implies that system polarization always occurs in the same location as an applied field.

However, given the non-zero size and hydrogen bonding structure of water and most

biomolecular species, this assumption is nearly always violated in biologically relevant systems





(Mobley, Barber, Fennell & Dill, 2008; Xie, Jiang, Brune & Scott, 2012 ). Finally, the static response assumes no time dependence for molecular polarization. However, even in bulk solvent, this time-independence is violated, with the optical dielectric constant of water of 2, which increases to the static value of 80 on the picosecond-nanosecond timescale (Fernandez, Mulev, Goodwin & Sengers, 1995; Zasetsky, 2011).

Given that nearly all of the assumptions of continuum electrostatics are violated in biologically relevant systems, the reader is probably wondering "why bother?" The answer lies in the power of heuristics. Although there are many arguments about its accuracy at microscopic levels (for example, (Schutz & Warshel, 2001; Kukic et al., 2013; Simonson, 2013)), the continuum model of water with a dielectric coefficient of 78-80 has proven remarkably useful for a wide range of applications. Likewise, while the *ab initio* derivation of solute dielectric constants is likely an exercise in futility, several heuristics have been useful in extending the applicability of continuum electrostatics to real biomolecular systems. These heuristics are described below; however, it is *essential* that the users of these continuum electrostatics heuristics are aware that they are using imperfect surrogates for complicated molecular phenomena. In particular, continuum electrostatics calculations should always be benchmarked for accuracy against real experimental data.

Simple – but imperfect – heuristics can be used to guide the selection of a molecular dielectric coefficient value ($\varepsilon_{solute}$). These are presented in Figure 1 and comprise three basic regimes:





- An $\varepsilon_{solute}$ of 2 represents the electronic polarization that will be found in any condensed matter system (Landau, Lifshiṱs & Pitaevskiĭ, 1984)]. This interpretation has an important implication for continuum electrostatics calculations: $\varepsilon_{solute} \geq 2$ should be used for all calculations with non-polarizable force fields (Leontyev & Stuchebrukhov, 2009).

- An $\varepsilon_{solute}$ of 4 has been ascribed to dried proteins and can be interpreted to include a very constrained polarization response of the protein dipoles (Gilson & Honig, 1986). This interpretation has an important implication for continuum electrostatics: an $\varepsilon_{solute} \geq 4$ should be used for all calculations that do not allow backbone rearrangement; e.g., through molecular dynamics or Monte Carlo configuration sampling.

- Larger values of $\varepsilon_{solute}$ allow more of the dipolar rearrangement of the backbone and side chains to be treated in an averaged manner with a single, compact parameter. Values of $4 < \varepsilon_{solute} < 12$ have been successfully used to predict protein-protein binding and are often attributed to limited side chain rearrangement. Values of $\varepsilon_{solute}$ above 12 are associated with larger scale backbone rearrangement and water penetration. Early continuum electrostatics attempts to model p$K_a$ values in proteins showed that $\varepsilon_{solute} = 20$ gave the best predictive power for calculations using a single dielectric constant and a single conformation (Antosiewicz, McCammon & Gilson, 1994).





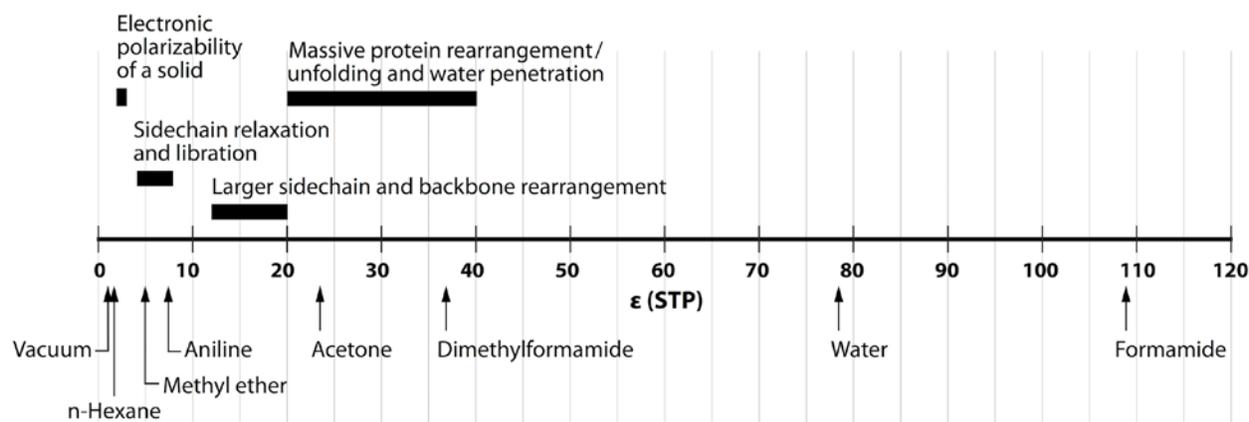

Figure 1.  A summary of dielectric constants of model compounds (bottom) and their application to protein continuum electrostatics modeling (top).

Not all continuum electrostatics treatments use a constant dielectric coefficient for the solute interior; some models use larger dielectric values for regions of the protein with greater responses to charge changes.  Alexov has varied the dielectric constant based on the atomic packing density (Li, Li, Zhang & Alexov, 2013).  While such variation does not explicitly take into account the chemical nature of the side chains, it does provide a mechanism for modeling internal degrees of freedom through the dielectric coefficient. Because these calculations avoid explicit conformational sampling, they offer the possibility of improved dielectric descriptions with the efficiency of standard continuum electrostatic methods.

Most continuum electrostatics software packages will identify interior cavities large enough to accommodate a water molecule – and many will assign these interior cavities a bulk dielectric value of $\varepsilon_{solute} = 80$. However, the high-dielectric treatment of internal cavities comes with a few important caveats.  First, it is difficult to provide a physical justification for a single





water molecule having the dielectric behavior of the bulk solvent. Second, this procedure is sensitive to small conformational changes that may cause regions to switch between $\varepsilon_{solute}$ and $\varepsilon_{solvent}$. To address this issue, Knapp has explored the effects of modeling the cavities with higher detail using a finer grid, which can accept smaller or less spherical wet regions, which improves the fit to benchmark $pK_a$s (Meyer, Kieseritzky & Knapp, 2011). Other methods make use of Gaussian dielectric boundaries in the calculation of the Poisson-Boltzmann equation, which also raises the effective internal dielectric constant (Word & Nicholls, 2011; Li, Li & Alexov, 2014).

As an alternative to high-dielectric models of internal cavities, continuum electrostatics software such as MCCE can include explicit water molecules within the protein (Song, Mao & Gunner, 2003). The included waters require explicit sampling. They must be optimized for each charge state and the number of waters may change with the charge state. Waters are often found in clusters so this optimization must be performed for multiple water molecules simultaneously. As a result, the inclusion of explicit water molecules can substantially increase the computational expense of the charge state calculation. The $pK_a$s obtained with implicit or explicit waters in the cavities have been found to agree surprisingly well in limited testing. .

The various modifications of the methods described above all improve the fit to known data essentially by increasing the effective interior dielectric constant. The electrostatic energy of a charge depends on the atomic charge distribution, the radius and the interior and exterior dielectric constant. Thus, the effective interior dielectric constant can be raised by increasing $\varepsilon_{solute}$ directly, or by smoothing the dielectric surface, or by enhancing cavities in the interior. The effects of changing these parameters have been explored separately. Without a better sense





of exactly how the various parameters interact the search through parameter space remains

Balkanized with different laboratories exploring their favorite parameters. However, it should be

noted that all of these changes do lead to significant improvement in the correspondence between

experimental and calculated values.

## **Modeling ion-solute interactions**

Ions are arguably more difficult to model than solvent. The simplest – and most widely

used – model of ion behavior is based on Debye-Hückel descriptions of aqueous ions as a diffuse

"cloud" that non-specifically screens electrostatic behavior in solution. The only major

determinants of ion behavior in Debye-Hückel-like models are the ion concentration and charge

valencies. However, this treatment has extreme limitations in describing realistic protein-ion

interactions that often include specific ion binding to protein sites as well as strong dependence

on ion species, even for ions with the same charge. To address these issues, some researchers

have begun to use models that combine implicit solvent descriptions with explicit simulation of

the ions (often via Monte Carlo sampling) (Sharp, Friedman, Misra, Hecht & Honig, 1995; Chen,

Marucho, Baker & Pappu, 2009; Song & Gunner, 2009). Nevertheless, many charge state

calculations still use Poisson-Boltzmann (PB) methods, which combine a Poisson treatment of

the solvent with the Boltzmann Debye-Hückel-like ion description.

## **Force field and parameter choices**

The microstate energy calculations described above require several different types of

parameters to describe molecular mechanics interactions, solvent characteristics, as well as





atomic size and charge. The molecular mechanics energies, atomic charges, and solute-solvent

Lennard-Jones interactions are often specified by standard molecular simulation force fields such

as AMOEBA (Schnieders, Baker, Ren & Ponder, 2007; Shi et al., 2013), AMBER (Pearlman et

al., 1995) or CHARMM (Brooks et al., 2009). These force fields can also be used to specify the

solute-solvent boundary through atomic radii; however, custom parameter sets such as PARSE

(Tannor et al., 1994) or ZAP (Word & Nicholls, 2011) are generally preferred because they have

been optimized to reproduce solvation energies.  In addition to atomic radii, the user must

specify the algorithm used to determine the shape of the solute-solvent interface.  A variety of

choices are available for these shape algorithms ranging from simple unions of spheres (Lee &

Richards, 1971) or Gaussians (Grant, Pickup, Sykes, Kitchen & Nicholls, 2007) to heuristic

molecular-accessible surfaces (Connolly, 1983) to thermodynamically defined self-consistent

solute-solvent interface definitions (Cheng, Dzubiella, McCammon & Li, 2007; Chen, Baker &

Wei, 2010, 2011).  Additionally, the user must choose a function to define the ion-accessible

regions around the protein; however, this interface is commonly chosen as an ion-accessible

union of spheres with radii equal to the atomic radii plus a nominal ionic radius of 0.2 nm.  It is

important to note that the optimal choices for radii, charges, and surface definitions are strongly

correlated; i.e., the radii are often optimized for a specific surface definition (Dong & Zhou,

2002).  These many choices of parameters are then presented to a program that will solve the

Poisson-Boltzmann equation to provide the solvation energy of individual conformers (within

the environment of the protein) and the pairwise interactions between all pairs of conformers.

For example, the programs DelPhi (Li, Li, Zhang & Alexov, 2012) or APBS (Baker, Sept,

Joseph, Holst & McCammon, 2001) have been employed within programs such as MCCE (Song





et al., 2009), and Karlsberg (Meyer et al., 2011), DelPhi p$K_a$ (Wang, Li & Alexov, 2015), and PDB2PKA (Dolinsky et al., 2007; Olsson, Sondergaard, Rostkowski & Jensen, 2011) to calculate the equilibrium protonation, redox and ligand binding states as a function of the appropriate chemical potential.

Titration state prediction methods must be benchmarked against datasets of *in situ* p$K_a$s, $E_m$s, or $K_d$s to determine their accuracy. There are approximately ≈350 wild-type residues with known p$K_a$s that are used extensively for such benchmarking (Song et al., 2009). These include a large number surface residues where the protein does not significantly influence the proton affinity. A subset of 100 residues has been selected to yield better range of p$K_a$s for training and testing (Stanton & Houk, 2008). The "null model" for charge state prediction assigns the model amino acid p$K_a$ value to all residues in the protein, regardless of their location or interactions. When the null model is used with the 100-residue subset, the RMSD between predicted and experimental p$K_a$ values is ≈1 pH unit. This sets a challenging metric for evaluating the performance of more sophisticated titration prediction methods. For example, the RMSD using modern Monte Carlo methods with continuum electrostatics force field, an $\varepsilon_{\text{solute}}$ between 4 and 8 and addition of conformational sampling or modification of the dielectric boundary and distribution can be between 0.9 and 1.1 (Song et al., 2009; Polydorides & Simonson, 2013; Wang et al., 2015). However, informatics-based methods such as PROPKA3 can do much better while sacrifice the underlying physical interactions for knowledge-based potentials (Olsson, 2011).

The Garcia-Moreno lab has placed >100 mutated residues into the core of Staphylococcal nuclease (Isom, Castaneda, Cannon, Velu & Garcia-Moreno, 2010; Isom, Castaneda, Cannon &





Garcia-Moreno, 2011; Richman, Majumdar & Garcia-Moreno, 2015). These residues formed the basis of the only blind challenge; i.e., where p$K_a$s were calculated without the experimental value being known (Nielsen et al., 2011). A meta-analysis (Gosink et al., 2014) of the blind predictions found that the RMSD for the null model is ≈3.5, indicating that the p$K_a$s for these residues were very shifted from the model values due to their burial in the protein. Empirical methods such as PROPKA3 (Olsson, 2011) did significantly better than the null model, methods with added conformational sampling did slightly better than the null model, while methods without added sampling did worse. Papers submitted after the p$K_a$s were revealed were able to obtain RMSDs <2 for this challenging dataset, as different modifications were explored once the errors were known. Particular improvement was found for methods that increased the response of the protein; e.g.: by using more explicit sampling via continuous-pH molecular dynamics (Wallace et al., 2011), by adding ensembles of structures obtained with MD (Witham et al., 2011), Rosetta (Song, 2011), through increased side-chain conformation sampling (Gunner, Zhu & Klein, 2011), by increasing the effective $\varepsilon_{solute}$ to implicitly model more internal water (Meyer et al., 2011), or by using a smoother dielectric boundary (Word & Nicholls, 2011). The errors for calculations with rigid backbones were smaller when crystal structures of the mutants were used rather then when the mutation was made *in silico*. Ensemble models which aggregated all of the predictions using Bayesian model averaging gave the best overall results (Gosink et al., 2014).





# **Conclusions**

The goal of this chapter was to present an overview of computational methods for predicting charge states of proteins with an emphasis on the issues that arise when applying continuum electrostatic methods to these applications. Given the many choices that must be made when applying these computational methods, one of the most important issues for this field is the availability of well-curated experimental data sets for testing computational predictions. The p$K_a$ Cooperative is a collaborative activity focused on assembling such data sets, performing blind predictions, and discussing the results as well as how to improve computational predictions (http://pkacoop.org/). All of the methods described above can be tuned to provide reasonable agreement with experimental data in a *post*diction setting. However, only a few methods perform with acceptable accuracy (~1 p$K_a$ unit error) in blind challenge predictions. Among these, constant-pH molecular dynamics methods generated the best predictions -- at significantly increased computational expense and the risk of poor convergence of the molecular dynamics simulations. Thus, computational methods continue to evolve to make the calculation of the energy of charges in protein faster and more accurate while providing increased physical insight into the forces at work. The current methods, despite their limitations, provide guidance as to the proton affinities of sites in proteins as well as the atomic interactions that affect a specific charge in a specific site and thus can be invaluable in getting more understanding of protein structure/function relationships.





# **Acknowledgements**

MRG gratefully acknowledges the support of Grant MCB1519640 from NSF, as well as National Institute on Minority Health and Health Disparities Grant 8G12MD7603-28 for infrastructure.  NAB gratefully acknowledges support from NIH grants R01GM069702 and R01GM099450.